\documentclass[prl,twocolumn]{revtex4}

\input epsf

\newcommand{\udar}{\updownarrow}
\newcommand{\lrar}{\leftrightarrow}
\newcommand{\ran}{\rangle}

\begin{document}

\title{Deterministic linear optics quantum computation \\
 utilizing linked photon circuits}

\author{N. Yoran and B. Reznik}
\affiliation{ School of Physics and Astronomy,
Raymond and Beverly Sackler Faculty of Exact Sciences, \\
Tel-Aviv University, Tel-Aviv 69978, Israel.}

\date{27 February 2003}

\begin{abstract}
We suggest an efficient scheme for quantum computation with linear
optical elements utilizing "linked" photon states. The linked
states are designed according to the particular quantum circuit
one wishes to process. Once a linked-state has been successfully
prepared, the computation is pursued {\em deterministically} by a
sequence of teleportation steps.  The present scheme enables a
significant reduction of the average number of elementary gates
per logical gate to about $20-30$ $CZ_{9/16}$ gates.
\end{abstract}


\maketitle



Optical systems have proven to be a very successful tool for
implementing quantum information and communication tasks such as
quantum cryptography, teleportation, and quantum dense
coding\cite{tasks}. However, when it comes to more complicated
protocols, let alone scalable quantum computation, such systems
suffer from a major disadvantage -- the lack of interaction
between photons that is needed for implementation of conditional
logic gates. In a recent work Knill, Laflamme and Milburn  (KLM)
\cite{KLM} proposed a scheme for quantum computation based on
linear optics, demonstrating that this obstacle can be overcome. A
two-qubit gate is performed, according to this scheme, in two
stages. First a standard ancillary state is prepared and subjected
to some gates. This off-line preparation step may succeed with low
probability. In the second stage the prepared state can be used to
apply a logical gate on the input state by means of the
teleportation scheme suggested by Gottesman and Chuang \cite{GC}.
This second step is again probabilistic, however, the
teleportation success probability can be made arbitrarily close to
unity, at the cost of a very significant increase of the required
number of elementary operations.

In this letter, we suggest a new scheme for quantum computation
with linear optical elements. A key point in our scheme is the
introduction of a multi-photon entangled "linked" photon state
whose structure is dictated by the form of the quantum circuit
that one wishes to construct. This state is prepared by employing
(KLM-type) non-deterministic gates. However, once the state has
been successfully constructed, the remaining computation process,
which utilizes a deterministic teleportation scheme, can be
ideally completed with unit probability. This
simplification results in a dramatic reduction in the required
number of elementary gate operation per logical gate.

The teleportation protocol employed here uses an extension of the
idea proposed by Popescu \cite{popescu} and experimentally
realized by De Martini {\it et al.} \cite{exp}. In this method a
pair of EPR entangled photons is utilized, and both the teleported
state and "half" of the EPR state are associated with a single
photon. Consequently, linear optics elements are sufficient for
implementing a complete Bell-state measurement. We begin by
extending this scheme to a chain of photons where each photon is
entangled with two nearest neighbors photons. Let us denote by
$p_1,p_2,...p_{n+1}$ the photons, and consider the following {\em
chain-state}
\begin{eqnarray}
 & & \chi_{p_1}\biggl(|1\ran_{p_{1}}|\udar\ran_{p_{2}}+
 |2\ran_{p_{1}}|\lrar\ran_{p_{2}}\biggr)\biggl(|3\ran_{p_{2}}
 |\udar\ran_{p_{3}}+|4\ran_{p_{2}}|\lrar\ran_{p_{3}}
 \biggr) \nonumber \\
 & &
 \cdots\biggl(|2n-1\ran_{p_{n}}|\udar\ran_{p_{n+1}}+|2n\ran_{p_{n}}
 |\lrar\ran_{p_{n+1}}\biggr)
|2n+1\ran_{p_{n+1}}
 \label{3}
\end{eqnarray}
$\chi_{p_1} =(\alpha|\udar\ran_{p_{1}}+\beta|\lrar\ran_{p_{1}})$
is an arbitrary polarization state of the first photon in the
chain, and $|m\ran$ and $|\udar,\lrar\ran$ denote the position and
polarization states of the photons, respectively. The chain state,
depicted schematically in Fig. 1, manifests pairwise maximal
entanglement, which we shall refer to as a $link$, between the
path of $p_i$ and the polarization of the next photon
$p_{i+1}$, and so forth.

\begin{center} \leavevmode
\epsfbox{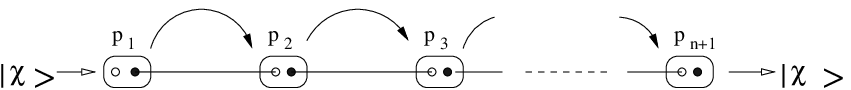}
\end{center}
\noindent {\small  FIG. 1. Schematic description of a single
chain-state with $n+1$ photons. A rectangle indicates a photon,
whose polarization and path degree of freedom are denoted by
empty and full circles, respectively.
  The linking horizontal lines denote the entanglement between
  position and polarization. The arrows portray the teleportation sequence
  of the state $\chi$ to the last photon in the chain. } \label{fig1}
\vspace{0.2cm}

The above state can be used to teleport the state
$|\chi\ran_{p_1}$ through the whole chain by means of $N$ separate
teleportation steps. In each step polarizing beam splitters (BPS)
are utilized for a path-polarization Bell measurement on photon
$p_i$. This sends the state $\chi$ (after correcting $p_{i+1}$) to
the polarization state of the next photon ($p_{i+1}$) in the
chain. By $N$ such sequential teleportation steps $|\chi\ran$ is
teleported to the last photon in the chain.

We shall use one such chain to represent the "world line" of a
single qubit in quantum circuit.  The time step corresponds here
to  teleportation steps. Hence for a circuit with $N$ input qubits
we will use $N$ chains. To include the required gates, we next
introduce gates between photons of different chains. Since the
teleportation of the input state sends $\chi$ to the polarization
states, we need to apply the gate between the polarizations states
of the appropriate photons in each chain. However as was shown by
Gottesman and Chuang \cite{GC} one can reverse the order of
teleportation and gate operations. Namely, we can first apply the
relevant gates on the different chains, producing a new state
(henceforth referred to as a linked-state) in which each photon in
a chain is entangled to another photon of a different chain, and
later teleport the input state through the linked-state.

To exemplify this construction, consider the three qubit circuit
depicted in Fig. 2, which sends $\chi_{1,2,3} \to
G_{1,2}G_{1,3}G_{1,2}\chi_{1,2,3}$. We replace this circuit with
the linked-state depicted in Fig. 3. that can be constructed as
follows. We begin with three chain-states,
$|I\ran=|p_1,p_2,p_3,p_4\ran_I$, $|II\ran=|p_1,p_2,p_3\ran_{II}$
and $|III\ran=|p_1,p_2\ran_{III}$. The gates are then applied on
the polarization states according to $|I,II,III\ran \to
G(I_{p_4},II_{p_3} ) G(I_{p_3},III_{p_2} ) G(I_{p_2},II_{p_2})
|I,II,III\ran$, where $I_{p_2}$ denotes the polarization states of
the second photon in the first chain, etc. Next, in order to
perform the computation, we introduce the input state $\chi$ by
rotating the polarization of the first photon in each chain
(assuming for the time being that $\chi$ is a non-entangled, known
state). Then we teleport $\chi$ through the linked-state and apply
the relevant single qubit corrections.

\begin{center} \leavevmode
\epsfbox{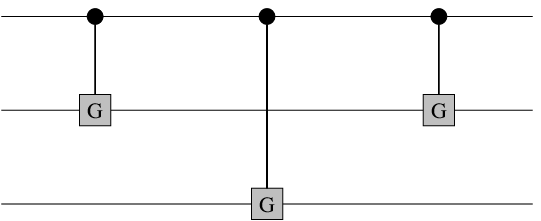}
\end{center}
\noindent{\small  FIG. 2. A simple circuit with three qubits and
gate operations.} \vspace{0.5cm}


\begin{center} \leavevmode
\epsfbox{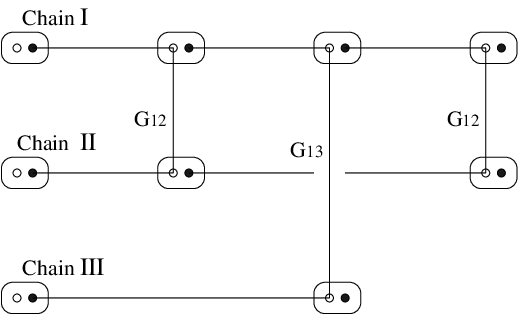}
\end{center}
\noindent {\small  FIG. 3. A schematic description of the linked
state that is needed for generating the quantum circuit in Fig. 2.
The vertical lines represent the entanglement that is produced by
applying the gates - G$_{i,j}$.}

\vspace {0.2cm}

Our scheme  generalizes to any quantum circuit, where the number
of links in each chain is determined by the number of gates that
are applied on that particular qubit.
We next address the preparation process in detail.

 {\em Preparation of linked states}. The off-line
part of the computation consists of two basic operations: addition
of a new link to each chain, and application of a two-qubit gate
between polarization states of different chains. We next show how
these two operations may be performed by applying KLM's
non-deterministic conditional phase flip $CZ$ gates. (We can use
either the basic or the improved gates proposed by KLM as well as
the improvements suggested in\cite{milburn,knill}). These gates
operate on the path degree of freedom of modes all with identical
polarization. This poses no difficulty in our case, because we can
easily move the information carried by polarization states back
and forth between the path and polarization degrees of freedom by
employing polarizing beam splitters and polarization rotation
plates.

Consider the construction of a new link to one of the chains (Fig.
4.). To achieve that, we apply a gate between the path degree of
freedom of the last photon in the chain and the polarization of an
additional photon. Suppose that $b$ is the last photon in the
chain $(|1\ran_{a}|\udar\ran_{b}+
|2\ran_{a}|\lrar\ran_{b})|3\ran_{b})$. We now add photon $c$ in a
state $(|5\ran_c + |6\ran_c)|\udar\ran_c$. Transmitting mode
$|3\ran_b$ through a $50/50$ beam splitters (splitting it to $3$
and $4$) and applying PBS we obtain
\begin{equation}
 |1\ran_{a}|\udar\ran_{b}(|3\ran_{b}+|4\ran_{b})+
 |2\ran_{a}|\lrar\ran_b(|3^{\prime}\ran_{b}+|4^{\prime}\ran_{b})
 \label{7}
\end{equation}
Notice that $3,4$ and $3',4'$ carry different polarizations, hence
we next apply $CZ$ in two consecutive steps. First between the
pair $\{3\,,\,4\}_b$ and $\{5\,,\,6\}_c$, followed by a $50/50$
beam splitter to $3$ and $4$. This will take the state of the four
modes to $|3\ran_b|5\ran_c+|4\ran_b|6\ran_c$. In the second step,
the polarization of modes $3^{\prime}$ and $4^{\prime}$ is rotated
(so it matches the polarization of $c$) and same the procedure is
repeated for $\{3^{\prime},4^{\prime}\}_b$. A successful sequence
of (two) gate operations creates  a link between $b$ and $c$.
Finally, the entanglement is transferred to the required
path-polarization form
\begin{equation}
 \cdots\times(|1\ran_{a}|\udar\ran_{b}+|2\ran_{a}|\lrar\ran_{b})\,
 (|3\ran_{b}|\udar\ran_{c}+|4\ran_{b}|\lrar\ran_{c})\,|5\ran_{c}
 \label{8}
\end{equation}

\begin{center} \leavevmode
\epsfbox{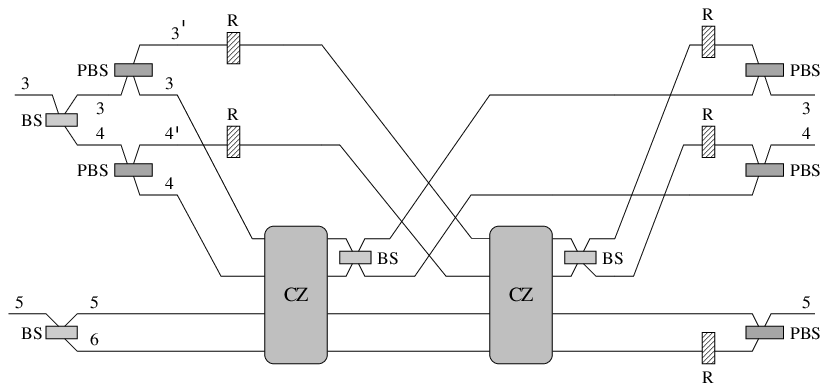}
\end{center}
\noindent {\small  FIG. 4. Addition of a link to a chain by the
 use of two $CZ$ gates. The additional elements
are $50/50$ beam splitters (BS), polarizing beam splitters
(PBS) and rotating wave plates (R).}
\label{fig5}
\vspace {0.2cm}

Having constructed the relevant links, we next consider a
two-qubit gate (Fig. 5).  In principle, the gate can be applied
after completing the construction of the chains. However, as we
apply the gate to the polarization of photons which are entangled
to other photons in the chain, this would require the application
of four $CZ$ gates for each logic gate. It would therefore be more
efficient to apply the gates to the proper photons immediately
after these links have been established, before the next links in
each chain are produced. As the last photon in each chain is
located in a single mode, the operation can be implemented by a
single $CZ$ application. Suppose that we want to apply a two-qubit
gate between the polarization states of photons $b$ and $d$ in the
chain states
\begin{eqnarray}
 \cdots\times(|1\ran_{a}|\udar\ran_{b}+|2\ran_{a}|\lrar\ran_{b})\,|5\ran_{b}&
 \quad \mbox{and} \nonumber \\
 \cdots\times(|3\ran_{c}|\udar\ran_{d}+|4\ran_{c}|\lrar\ran_{d})\,|6\ran_{d}&
 \label{4}
\end{eqnarray}
Employing a PBS for each chain and rotating the polarization of the
modes corresponding to the horizontal modes ($5^{\prime}$ and
$6^{\prime}$) we obtain
\begin{eqnarray}
 \cdots\times|\udar\ran_{b}\,(|1\ran_{a}|5\ran_{b}\,+\,
 |2\ran_{a}|5^{\prime}\ran_{b})&
 \quad \mbox{and} \nonumber \\
 \cdots\times|\udar\ran_{d}\,(|3\ran_{c}|6\ran_{d}\,+\,
 |4\ran_{c}|6^{\prime}\ran_{b})&  \label{5}
\end{eqnarray}
At this point we can apply the $CZ$ gate which produces a
conditional phase flip (other two qubit gates are equivalent up to
singe-qubit operations). Finally we transfer the entanglement
between photons in each pair $\{a\,,\,b\}$ and $\{c\,,\,d\}$ back
to the path polarization form. A successful $CZ$ gate operation
this leaves us in the desired state:
\begin{eqnarray}
 (\:|1\ran_{a}|\udar\ran_{b}|3\ran_{c}|\udar\ran_{d}\,
 +\,|1\ran_{a}|\udar\ran_{b}|4_{c}\ran|\lrar\ran_{d}\,+&
 \nonumber \\ |2\ran_{a}|\lrar\ran_{b}|3\ran_{c}|\udar\ran_{d}
 \,-\, |2\ran_{a}|\lrar\ran_{b}|4\ran_{c}|\lrar\ran_{d}\,)&\:
 |5\ran_{b}|6\ran_{d}
 \label{6}
\end{eqnarray}

\begin{center} \leavevmode
\epsfbox{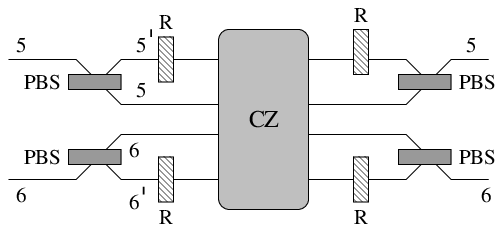}
\end{center}
\noindent {\small  FIG. 5. Conditional phase flip on two photons
of
  two different chains with a single $CZ$ operation.}
\label{fig4}

{\em Efficient construction of large circuits.} A basic building
block of the KLM scheme is a conditional phase flip gate that
employs interference of input photons and ancillary photons and
post-selection. This basic gate operates successfully with
probability $1/16$. In our scheme this gate can be used for
generating small circuits. However, for long enough quantum
circuits the preparation process becomes inefficient, and gates
with higher success rate must be employed. In the construction of
the overall linked-state we proceed step by step, since a failure
in the gate operation in one step might destroy previously
constructed links and gates, we require that the combined two
stage process of link/gate generation has on average a probability
larger than $1/2$ to progress. This can be achieved by replacing
the basic $CZ$ gates with an improved gate version proposed by KLM
(for a recent improvement see \cite{franson}). These gates operate
through the application of a new type of teleportation protocol
based on the $n+1$ point Fourier transform ($\hat{F}_{n+1}$),
which operates successfully with probability $n/(n+1)$. This gate,
$CZ_{n^{2}/(n+1)^{2}}$, is constructed of two independent
$\hat{F}_{n+1}$-based teleportations and therefore operates
successfully with probability $n^{2}/(n+1)^{2}$. The application
of each $CZ_{n^{2}/(n+1)^{2}}$ requires that a special ancillary
state of $2n$ photons in $4n$ modes would be prepared in advance
(by utilizing basic $CZ$ gates). Thus, the preparation stage of
our scheme has two parts. In the first part we prepare independent
small-scale ancillary states with which we apply, in the second
part, the $CZ_{n^{2}/(n+1)^{2}}$ gates to construct the overall
linked-state.

The $CZ_{n^{2}/(n+1)^{2}}$ fails when either one of the
independent teleportation protocols fails. A failure of the
teleportation protocol results in the measurement of the
teleported qubit. Let us consider the operation of applying a gate
between two chains. By inspecting Eq. (\ref{5}), it is clear that
failure in teleporting photon $b$ would leave photon $a$ in either
mode $1$ or mode $2$ breaking the link. In the same way failure in
teleporting photon $d$ would brake its corresponding link.
Clearly, it would be more efficient to apply the two teleportation
protocols in a sequence, where the second is applied only if the
first has succeeded, eliminating the possibility of breaking two
links. Thus, in applying a gate, the probability of success is
$p=n^{2}/(n+1)^2$ while with probability $(1-p)$ one link is
broken.

In adding a link to a chain we apply two $CZ$ operations. These
are applied to one photon (in four modes) which constitutes the
last link of a chain, together with a newly introduced photon. As
this new photon is not entangled to any chain, there is no point
in wasting an $\hat{F}_{n+1}$-based teleportation protocol on it.
This photon can be prepared as part of the ancilla. Therefore,
each of the $CZ$ operations in adding a link will be carried out
through the application of a "half" $CZ_{n^{2}/(n+1)^{2}}$ gate in
which a pair of modes ($3$ and $4$ in~(\ref{7}) and afterwards
$3^{\prime}$ and $4^{\prime}$) undergoes teleportation, but the
other pair ($5$ and $6$) does not. The $CZ$ is applied to this
pair of modes together with the components of the ancilla in the
first part of the preparation. The ancillary state \cite{KLM} for
one $\hat{F}_{n+1}$-based teleportation is $|t_{n}\rangle =
\sum_{j=0}^{n}|1\ran^{j}|0\ran^{n-j}|0\ran^{j}|1\ran^{n-j}$,
defining a modified $|t_{n}\ran$ as $ |\tilde{t}_{n}\ran =
\sum_{j=0}^{n} (-1)^{j}
 |1\ran^{j}|0\ran^{n-j}|0\ran^{j}|1\ran^{n-j}$
The ancillary state for the two $CZ$ operations would be
\begin{equation}
 |5\ran|\tilde{t}_{n}\ran_{1}|\tilde{t}_{n}\ran_{2}
 \:+\:|6\ran|t_{n}\ran_{1}|t_{n}\ran_{2} \label{71}
\end{equation}
where $1$ and $2$ denote the states used for the teleportation of
modes $\{3,\;4\}$ and $\{3^{\prime},\;4^{\prime}\}$ respectively.
In this case, the overall process can fail in two ways. If a
maximal number of photons is detected at the outputs of the
$\hat{F}_{n+1}$ operation (thus destroying also the teleported
photon), then the previous link is destroyed together with the
gate operation that was applied to it. If no photon is detected
the entanglement is not completely destroyed and we can still
bring the system back to the initial state. Thus, the whole
process (the two teleportation protocols) succeeds with
probability $p=n^{2}/(n+1)^{2}$ and in the case of failure we have
the same probability ($(1-p)/2$)to either destroy the previous
link (together with the gate operation) or to remain in the same
initial state.


Employing our scheme for every logic gate in the required
computation we need to perform six successful $\hat{F}_{n+1}$
based teleportation protocols, which are equivalent to three
$CZ_{n^{2}/(n+1)^{2}}$ gates (two for adding two links to two
different chains and one for applying the gate to those links).
$n=3$  is  the smallest value for which the step-by-step
construction of the linked-state can advance forward with higher
probability than moving backwards. In a computer simulation of a
construction process for a two qubit linked-state (a two qubit
circuit), we obtained the average number of gates per logical
gate,  of ~$\sim220$  for a $CZ_{9/16}$ gate, and ~$\sim15$ for a
 $CZ_{16/25}$ gate.
 The large average number required for the case of the $CZ_{9/16}$,
results from an overall probability very close to $1/2 $ to
advance forward. This number can be however significantly reduced
if we change the overall construction by adding inert links to the
chains, i.e. photons on which no gate operation is applied. In
this case, for a qubit that takes part in $n$ two-qubit gates we
construct a chain of $2n$ links while the gates are applied to
every second link. The only purpose of the inert links is to
prevent a possible failure from spreading backwards, destroying
previously constructed links and gates (if the step of adding a
second link fails destructively then only the previous link is
destroyed, the previous gate is not affected). Using this type of
construction we need five successful gate operations for every
logic gate. Applying a computer simulation on a pair of chains we
obtain the number of $\sim23$ $CZ_{9/16}$ applications on average
for every logic gate. The gate $CZ_{4/9}$ can also be used by
introducing additional inert links (at least three). When six
inert links are added for each logic gate (thus, requiring $13$
successful gate operations for each logic gate) we get around
$\sim220$ $CZ_{4/9}$ applications on average per gate.

It should be noted that for large scale linked-states we can start
the computation before the completion of the preparation. The
computation is then carried out simultaneously with the
construction of the liked state, where a 'safety margin' (in terms
of the number of links) is maintained, keeping the probability for
a sequence of failures that could destroy part of the data
negligible. This method is more economical in terms of the
required quantum storage capacity, as only a part of the complete
linked-state is kept at a given instance \cite{Knill2}.

So far we have assumed that the input is received classically. If
we receive a quantum state as an input, where each qubit is
encoded for example in two modes, then by measuring this qubit
together with the path degree of freedom of the photon in the
first link in the Bell basis will transfer this qubit to the head
of the chain (to the polarization of the second photon as the
first was measured). This measurement can be accomplished by
applying a $CZ$ gate together with additional one-qubit
operations. Clearly, as this operation involves the actual input,
the applied $CZ$ gate must be one with a very high success rate.
However, this operation is carried out only once for each qubit.

In the present scheme each photon carries two qubits. To achieve
that we utilized, both polarization and path degree of freedom.
This led to a simple path-polarization factorization of the linked
state (1). However, our scheme does not require the use of
polarization. It has been shown that a linear optical realization
exists for any $N\times N$ unitary matrix \cite{Reck}. We can
therefore represent the chain state (1) in terms of path degrees
of freedom alone, by attributing four possible modes to each
photon.

We thank Lev Vaidman for very helpful discussions, and acknowledge
support from the Israel Science Foundation (grant 62/01-1) , and
MOD Research and Technology Unit.


\end{document}